\documentclass[aps, prd, showpacs, showkeys, floatfix, superscriptaddress, nofootinbib, longbibliography]{revtex4-2}

\usepackage{graphicx}

\def\vp{\varphi}
\def\ra{\rightarrow}
\def\ds{d^s\!x}
\def\half{\textstyle{\frac{1}{2}}}
\def\tint{{\textstyle\int}}

\def\ra{\rightarrow}

\def\k{\kappa}

\def\b{\begin{eqnarray*}}  %takes no eqn numbers
\def\e{\end{eqnarray*}}    %takes no eqn numbers
\def\bn{\begin{eqnarray}}  %takes eqn numbers
\def\en{\end{eqnarray}}    %takes eqn numbers

\begin{document} 
%%%%%%%%%%%%%%%%%%%%%%%%%%%%%%%%%%%%%%%%%%%%%%%%%%%%%%%%%%%%%%%%%%%%%%%%%%%%%%
%%%%%%%%%%%%%%%%%%%%%%%%%%%%%%%%%%%%%%%%%%%%%%%%%%%%%%%%%%%%%%%%%%%%%%%%%%%%%%
%%%%%%%%%%%%%%%%%%%%%%%%%%%%%%%%%%%%%%%%%%%%%%%%%%%%%%%%%%%%%%%%%%%%%%%%%%%%%%
\title{Scaled Affine Quantization of Ultralocal $\vp^4_2$ a comparative Path Integral Monte 
Carlo study with scaled Canonical Quantization}

\author{Riccardo Fantoni}
\email{riccardo.fantoni@posta.istruzione.it}
\affiliation{Universit\`a di Trieste, Dipartimento di Fisica, \\
strada Costiera 11, 34151 Grignano (Trieste), Italy}
\author{John R. Klauder}
\email{klauder@ufl.edu}
\affiliation{Department of Physics and Department of Mathematics \\
University of Florida,   %P.O. Box 118440\\
Gainesville, FL 32611-8440}

\date{\today}

\begin{abstract}
After the success of affine quantization in proving through Monte Carlo analysis 
that the covariant euclidean scalar field theory, $\vp^r_n$, where $r$ denotes the 
power of the interaction term and $n = s + 1$ with $s$ the spatial dimension and 
$1$ adds imaginary time, such that $r \geq 2n/(n-2)$ can be acceptably quantized and the 
resulting theory is nontrivial, unlike what happens using canonical quantization, we show 
here that the same has to be expected for $r>2$ and any $n$ even for the ultralocal field 
theory. In particular we consider the ultralocal $\vp^4_2$ model and study its renormalized 
properties for both the scaled canonical quantization version and the scaled affine 
quantization version through path integral Monte Carlo.
\end{abstract}

\maketitle
%%%%%%%%%%%%%%%%%%%%%%%%%%%%%%%%%%%%%%%%%%%%%%%%%%%%%%%%%%%%%%%%%%%%%%%%%%%%%%
\section{Introduction}
%%%%%%%%%%%%%%%%%%%%%%%%%%%%%%%%%%%%%%%%%%%%%%%%%%%%%%%%%%%%%%%%%%%%%%%%%%%%%%
Ultralocal euclidean scalar field quantization, henceforth denoted $\vp^r_n$, where $r$ 
is the power of the interaction term and $n = s + 1$ where $s$ is the spatial 
dimension and $1$ adds imaginary time, such that $r < 2$ can be treated by 
canonical quantization (CQ), while models such that $r > 2$ and any $n\geq 2$ are trivial 
\cite{Klauder2020b,Freedman1982,Aizenman1981,Frohlich1982,Siefert2014}. However, 
there exists a different approach called {\sl affine quantization} (AQ) 
\cite{Klauder2020b,Klauder2000,Klauder2020c} that promotes a different set 
of classical variables to become the basic quantum operators and it offers different 
results, such as models for which $r > 2$. In particular one can show that while the 
Fubini-Study metric for the canonical coherent states that evaluates the distance-squared 
between two infinitesimally close ray-vectors (minimized over any simple phase) leads to
a {\sl flat space} that already involves Cartesian coordinates, in the affine case the 
Fubini-Study metric describes a {\sl Poincar\'e half plane} \cite{Klauder2020a,Klauder2020c}, 
has a constant negative curvature \cite{Fantoni2003}, and is geodesically complete.
Unlike a flat plane, or a constant positive curvature surface (which holds the metric of
three-dimensional spin coherent states), a space of constant negative curvature can not be 
visualized in a 3-dimensional flat space. At every point in this space the negative curvature 
appears like a saddle having an ``up curve'' in the direction of the rider’s chest and a 
``down curve'' in the direction of the rider’s legs. 
\footnote{Of course other types of more complex quantizations may still be possible who 
involve non-constant curvature surfaces \cite{Fantoni08c,Fantoni19a}.}

In the present work we show, with the aid of a Path Integral Monte Carlo (PIMC) analysis, 
that $r = 4$ and $n=2$ can be acceptably quantized using scaled affine quantization which had 
been previously successfully used for the covariant case \cite{Fantoni2020,Fantoni2020a,Fantoni2020b,Fantoni2021,Fantoni2021b,Fantoni2022,Fantoni2022b,Fantoni2022c} 

Being the current study in a lower, therefore unphysical, spacial dimension it nonetheless 
allowed us to get closer to the continuum limit than it was feasible for the physically 
relevant four-dimensional case on the computer due to the rapid increase of necessary lattice 
points as dimensionality is increased. Therefore this work can indirectly give us a better 
understanding of the physically relevant case which had been already preliminarily studied by 
us in its covariant version \cite{Fantoni2020a}. Interestingly, the triviality of the scaling 
limits of the canonical Ising and covariant self-interacting scalar field models in four 
dimensions has been rigorously demonstrated recently \cite{Aizeman2021}.

%%%%%%%%%%%%%%%%%%%%%%%%%%%%%%%%%%%%%%%%%%%%%%%%%%%%%%%%%%%%%%%%%%%%%%%%%%%%%%
\section{A Comparison Between Canonical Quantization and Affine Quantization for 
ultralocal Fields}
%%%%%%%%%%%%%%%%%%%%%%%%%%%%%%%%%%%%%%%%%%%%%%%%%%%%%%%%%%%%%%%%%%%%%%%%%%%%%%

%%%%%%%%%%%%%%%%%%%%%%%%%%%%%%%%%%%%%%%%%%%%%%%%%%%%%%%%%%%%%%%%%%%%%%%%%%%%%%
\subsection{Canonical quantization (CQ) of scalar fields}
%%%%%%%%%%%%%%%%%%%%%%%%%%%%%%%%%%%%%%%%%%%%%%%%%%%%%%%%%%%%%%%%%%%%%%%%%%%%%%
Let us begin with the classical Hamiltonian for a single ultralocal field $\vp(x)$
\bn H(\pi,\vp) = \int \{ \half[ \pi(x)^2+m^2\,\vp(x)^2]+g \,|\vp(x)|^r
\;\}\;\ds\;, \en
where $n=s+1$ is the number of spacetime variables, and $r$ is any real number.
When $g$ is zero, the remaining expression involves a domain in which a full set of 
variables, i.e., $\pi(x)$ and $\vp(x)$, lead to a finite Hamiltonian value. If $g=0\ra g>0$, 
there are two possible results. If $r<2$, then the domain remains the same. However, if 
$r>2$, then there is a new domain that is smaller than the original domain because the 
interaction term $\tint |\vp(x)|^r\;\ds =\infty$ leads to a reduction of certain fields. 
Since we work in a finite volume region, the fields that cause that divergence are not 
$\vp(x)=\infty$, because that would have eliminated the original domain when $g=0$. The only 
way for $\tint |\vp(x)|^r\;\ds=\infty$ is, for example, given by $\vp(x)=1/|x-c|^k$ 
where $k$ is small enough so that $\tint \vp(x)^2\;\ds<\infty$, while $r>2$ is big enough so 
that $\tint |\vp(x)|^r\;\ds =\infty$ for example for $r=5/2$. Such behavior leads to 
immediate results in perturbation infinities in a power series of $g$, leading to a 
non-renormalizable process, for which quantum efforts, using canonical quantization, collapse 
to ``free'' results, despite that $g>0$, as all that is continuously connected to the 
original free theory where $g=0$.
Here we should be more precise since the relevant quantity to look at is the action rather 
than the Hamiltonian, so we should really compare the interaction term 
$\tint |\vp(x)|^r\;d^nx$ and the kinetic term $\tint [\partial\vp(x)/\partial x_0]^2\;d^nx$ 
or the mass term $\tint \vp(x)^2\;d^nx$. 
If we consider stationary fields as particular cases then the relevant integral is the mass 
term and we immediately see that we may have triviality for $r>2$. But the same remains true 
also for space independent fields.
\footnote{If we consider $\vp(x)=1/|x_0-c|^k$ then in order to have 
${\cal D}_{g>0}\subset {\cal D}_{g=0}$, where ${\cal D}_{g}$ is the domain of those 
$\varphi(x)$, in the ultralocal theory, where the action is not divergent, we require a 
divergent interaction term but a convergent kinetic term or $rk>1>2(k+1)$ that is possible if 
$r>-2$. Additionally, since we always want a convergent mass term we must also have 
$0<k<1/2$ which again requires $r>2$ for triaviality. This means that for $r>2$ the 
domains change because of reducing $g$ back to zero will only retain the smallest version of 
the domain by continuity, and that will not be the theory you started out with. For space 
independent fields the ultralocal theory is the same as the covariant theory which is 
trivial for $r>2n/(n-2)>2$. This is due to the fact that 
${\cal D}_g^{\rm covariant}\neq{\cal D}_g^{\rm ultralocal}$.}  
As we will see in Section \ref{sec:MC}, our numerical results give evidence for a 
``free'' behavior of the CQ theory in this case. 
   
Having seen what CQ can show us, now let us turn to AQ.

%%%%%%%%%%%%%%%%%%%%%%%%%%%%%%%%%%%%%%%%%%%%%%%%%%%%%%%%%%%%%%%%%%%%%%%%%%%%%%
\subsection{Affine quantization (AQ) of scalar fields}
%%%%%%%%%%%%%%%%%%%%%%%%%%%%%%%%%%%%%%%%%%%%%%%%%%%%%%%%%%%%%%%%%%%%%%%%%%%%%%
The classical affine variables are the {\sl dilation} $\k(x)\equiv \pi(x)\,\vp(x)$ and the 
field $\vp(x)\neq 0$. The reason we insist that $\vp(x)\neq 0$ is because if $\vp(x)=0$ and 
$\k(x)=0$ then $\pi(x)$ is not well defined.
   
We next introduce the same classical Hamiltonian we chose before now expressed in affine 
variables. This leads us to
\bn H'(\k,\vp) =\int\{\half[\k(x)^2\,\vp(x)^{-2}+m^2\,\vp(x)^2]+g\,
|\vp(x)|^r\}\;\ds \;, \en
in  which $\vp(x)\neq 0$ is an important fact. With these variables we see that $\pi=k/\vp$ 
so we should not let neither $\vp=0$ nor $\vp=\pm\infty$ otherwise in either cases we could 
find a form of indecision ($0/0$ or $\infty/\infty$) for the dilation $k$ which would then be 
not well defined. The essential result $0<|\vp(x)|<\infty$, leads to the fact that these AQ 
bounds on $\vp(x)$  {\it forbid any non-renormalizability}, a `disease' which plagues the CQ 
analysis.  With AQ, this new insight implies that any model $\vp^r_n$ does not become a 
``free'' result, but leads to an appropriate ``non-free'' result. 
  
What follows in the coming sections is additional PIMC studies using AQ and CQ procedures. 

%%%%%%%%%%%%%%%%%%%%%%%%%%%%%%%%%%%%%%%%%%%%%%%%%%%%%%%%%%%%%%%%%%%%%%%%%%%%%%
\section{Lattice formulation of the field theory}
%%%%%%%%%%%%%%%%%%%%%%%%%%%%%%%%%%%%%%%%%%%%%%%%%%%%%%%%%%%%%%%%%%%%%%%%%%%%%%
The quantum affine operators are the scalar field $\hat{\vp}(x)=\vp(x)$ and the 
{\it dilation} operator 
\footnote{Since $\vp(x)\neq 0$, that means $\pi^\dagger \neq \pi$ so, to make that clear 
we should say that 
$\hat{\k}(x)\equiv[\hat{\vp}(x)\hat{\pi}(x)+\hat{\pi}^\dagger(x)\hat{\vp}(x)]/2$ 
to make sure that $\hat{k}^\dagger=\hat{k}$. But 
$\hat{\pi}^\dagger\hat{\vp}=\hat{\pi}\hat{\vp}$ because in that case 
$\hat{\pi}^\dagger$ acts like $\hat{\pi}$ thanks to having $\hat{\pi}$ acting on 
$\hat{\vp}$.}
$\hat{\k}(x)=[\hat{\vp}(x)\hat{\pi}(x)+\hat{\pi}(x)\hat{\vp}(x)]/2$ 
where the momentum operator is $\hat{\pi}(x)=-i\hbar\delta/\delta\vp(x)$. Accordingly for 
the self adjoint kinetic term 
$\hat{\k}(x)\hat{\vp}(x)^{-2}\hat{\k}(x)=\hat{\pi}(x)^2+(3/4)\hbar\delta(0)^{2s}\vp(x)^{-2}$ 
and one finds for the quantum Hamiltonian operator
\bn \label{eq:HO}
\hat{H'}(\hat{\k},\hat{\vp}) =\int\left\{\half[\hat{\pi}(x)^2+m^2\,\vp(x)^2]+g\,|\vp(x)|^r +{\textstyle\frac{3}{8}}\hbar^2\frac{\delta(0)^{2s}}{\vp(x)^2}\right\}\;\ds.
\en

As in previous works \cite{Fantoni2021b,Fantoni2022,Fantoni2022c} we use the scaling 
$\pi\ra a^{-s/2}\pi, \vp\ra a^{-s/2}\vp, g\ra a^{s(r-2)/2} g$, which is 
necessary 
\footnote{\label{foot:math-phys} Note that from a physical point of view one never has to 
worry about the mathematical divergence since the lattice spacing will necessarily have a 
lower bound. For example at an atomic level one will have $a\gtrsim 1$\AA. In other words the 
continuum limit will never be a mathematical one.}
to eliminate the Dirac delta factor $\delta(0)=a^{-1}$ divergent in the continuum limit 
$a\to 0$. Of course for $r>2$ the rescaled coupling constant, $g$, vanishes in the continuum 
limit since $a\to 0$, therefore we expect no difference between the interacting and the free 
model in such a limit. The theory considers a real scalar field $\vp$ taking the value 
$\vp(x)$ on each site of a periodic, hypercubic, $n$-dimensional lattice of lattice spacing $a$, our ultraviolet cutoff, and periodicity $L=Na$. The affine action for the field, 
${\cal S}'=\int \bar{H'}\,dx_0$ 
(with $x_0=ct$ where $c$ is the speed of light constant and $t$ is imaginary time), with 
$\bar{H'}$ the semi-classical Hamiltonian corresponding to the one of Eq. (\ref{eq:HO}), is 
then approximated by
\bn \label{eq:scaled-affine-action} \nonumber
{\cal S}'[\vp]/a&\approx&\half\left\{\sum_{x}a^{-2}[\vp(x)-\vp(x+e_0)]^2 
+m^2\sum_{x}\vp(x)^2\right\}+\sum_{x}g\,|\vp(x)|^r\\
&&+{\textstyle\frac{3}{8}\sum_{x}}\hbar^2{\displaystyle\frac{1}{\vp(x)^2}},
\en
where $e_\mu$ is a vector of length $a$ in the $+\mu$ direction. Respect to the previously 
considered covariant case \cite{Fantoni2020,Fantoni2020a,Fantoni2020b,Fantoni2021,Fantoni2021b,Fantoni2022,Fantoni2022b,Fantoni2022c}, 
being now absent derivatives with respect to space, the field is allowed to be 
discontinuous in space (but will still be continuous in time).

The corresponding canonical action, ${\cal S}=\int \bar{H}\,dx_0$, is then approximated by
\bn \label{eq:canonical-action}
{\cal S}[\vp]/a\approx&\half\left\{\sum_{x}a^{-2}[\vp(x)-\vp(x+e_0)]^2 
+m^2\sum_{x}\vp(x)^2\right\}+\sum_{x}g\,|\vp(x)|^r.
\en

In this work we are interested in reaching the continuum limit by taking $Na$ fixed 
and letting $N\to\infty$ at fixed volume $L^s$ and absolute temperature $T=1/k_BL$ with 
$k_B$ the Boltzmann's constant.

The vacuum expectation value of an observable ${\cal O}[\vp]$ will then be given by the 
following expression
\bn \label{eq:EV}
\langle{\cal O}\rangle=\frac{\int{\cal O}[\vp]\exp(-{\cal S}[\vp])\;{\cal D}\vp(x)}{\int\exp(-{\cal S}[\vp])\;{\cal D}\vp(x)},
\en
where the functional integrals will be calculated on a lattice using the path integral 
Monte Carlo method as explained further on.

%%%%%%%%%%%%%%%%%%%%%%%%%%%%%%%%%%%%%%%%%%%%%%%%%%%%%%%%%%%%%%%%%%%%%%%%%%%%%%
\section{Path Integral Monte Carlo (PIMC) simulation}
%%%%%%%%%%%%%%%%%%%%%%%%%%%%%%%%%%%%%%%%%%%%%%%%%%%%%%%%%%%%%%%%%%%%%%%%%%%%%%
\label{sec:MC}
We performed PIMC \cite{Metropolis,Kalos-Whitlock,Ceperley1995,Fantoni12d} for the action of 
Eq. (\ref{eq:scaled-affine-action}) with $r=4$ and $n=2$. In particular we calculated the 
renormalized coupling constant $g_R$ and mass $m_R$ defined in Eqs. (11) and (13) of 
\cite{Fantoni2020a} respectively, measuring them in the path integral MC through vacuum 
expectation values like in Eq. (\ref{eq:EV}). In particular 
$m_R^2=p_0^2\langle|\tilde{\vp}(p_0)|^2\rangle/[\langle\tilde{\vp}(0)^2\rangle-\langle|\tilde{\vp}(p_0)|^2\rangle]$ and at zero momentum
$g_R=[3\langle\tilde{\vp}(0)^2\rangle^2-\langle\tilde{\vp}(0)^4\rangle]/\langle\tilde{\vp}(0)^2\rangle^2$, where $\tilde{\vp}(p)=\int d^nx\;e^{ip\cdot x}\vp(x)$ is the 
Fourier transform of the field and we choose the 2-momentum $p_0$ with the zero
component equal to $2\pi/Na$ and the other component equal to zero. Since the integration 
variables in Eq. (\ref{eq:EV}) are 
$N^n$, being able to choose $n=2$ allowed us to greatly speed up the calculations compared to 
our previous covariant studies for $n>2$ \cite{Fantoni2020,Fantoni2020a,Fantoni2020b,Fantoni2021,Fantoni2021b,Fantoni2022,Fantoni2022b,Fantoni2022c} 
and this made possible to push ourselves closer to the continuum limit, to bigger $N$.

\begin{figure}[htbp]
\begin{center}
\includegraphics[width=7cm]{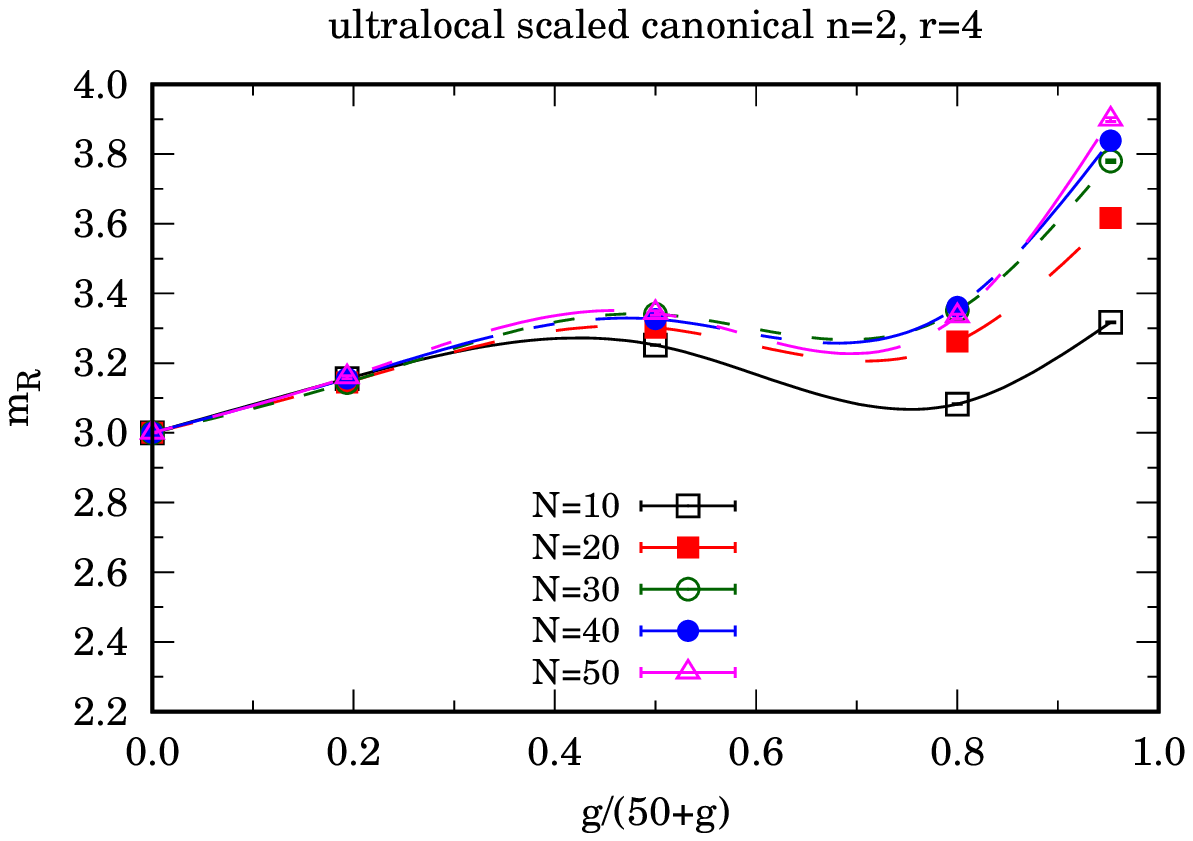}
\includegraphics[width=7cm]{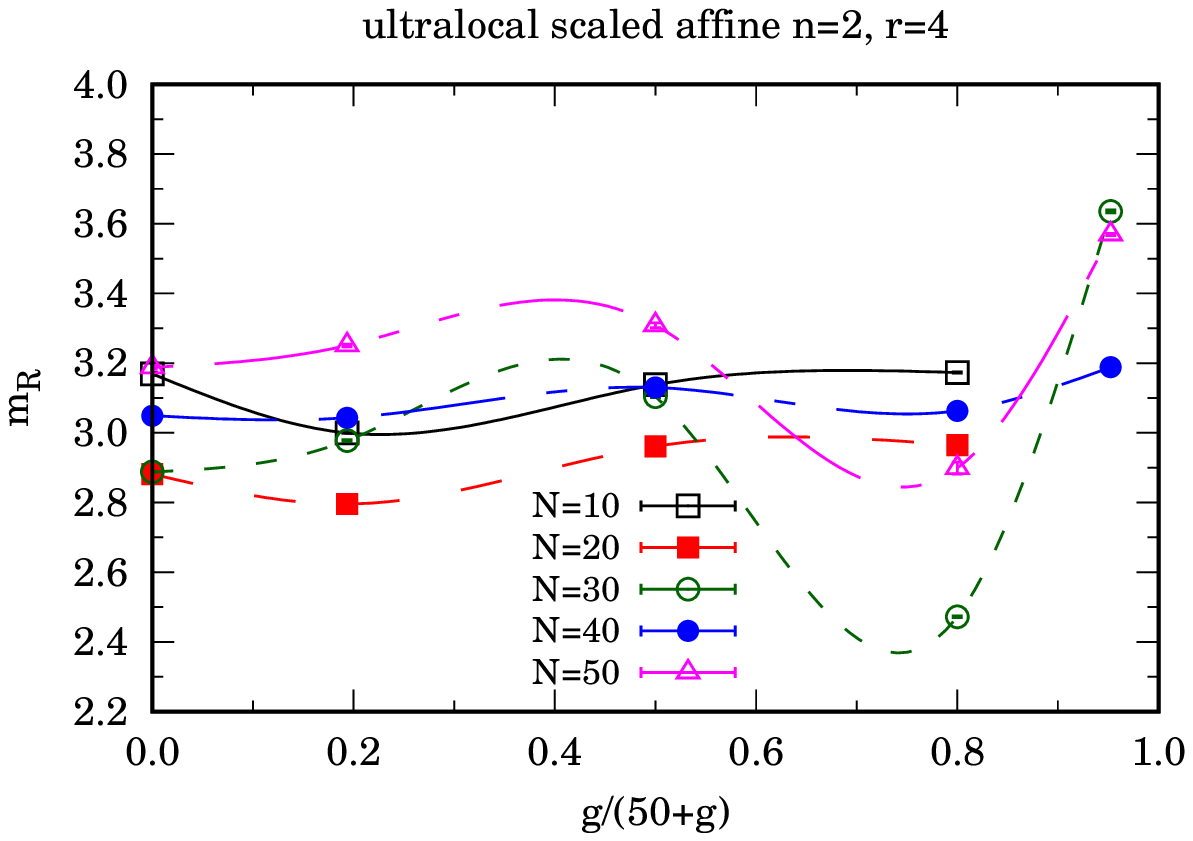}
\\
\includegraphics[width=7cm]{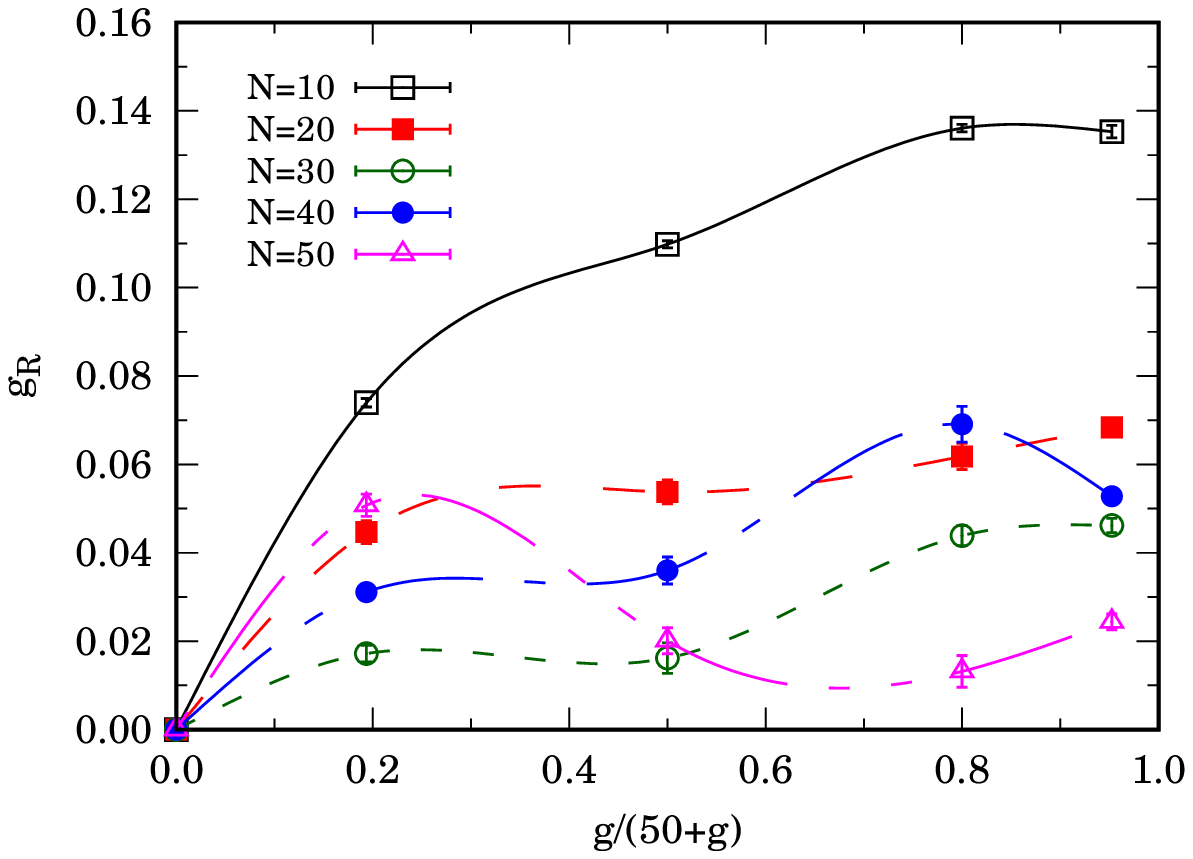}
\includegraphics[width=7cm]{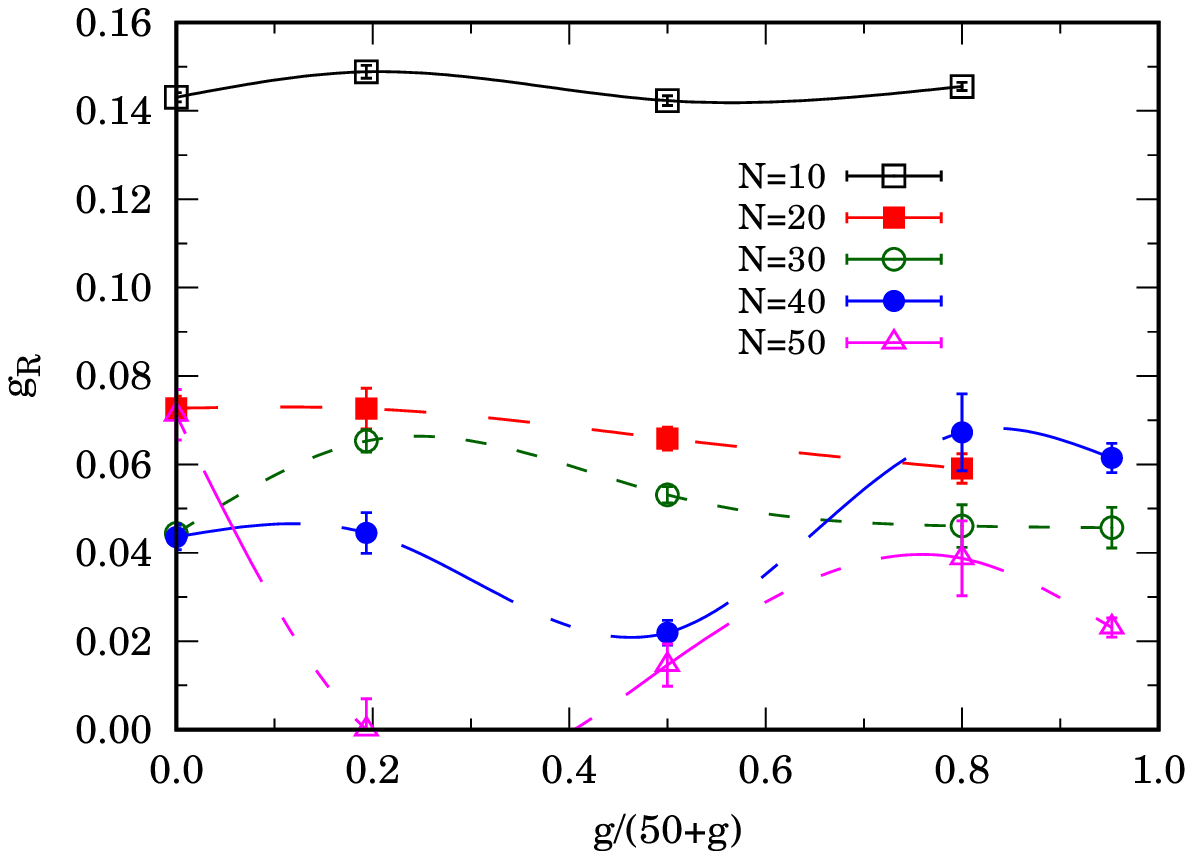}
\\
\includegraphics[width=7cm]{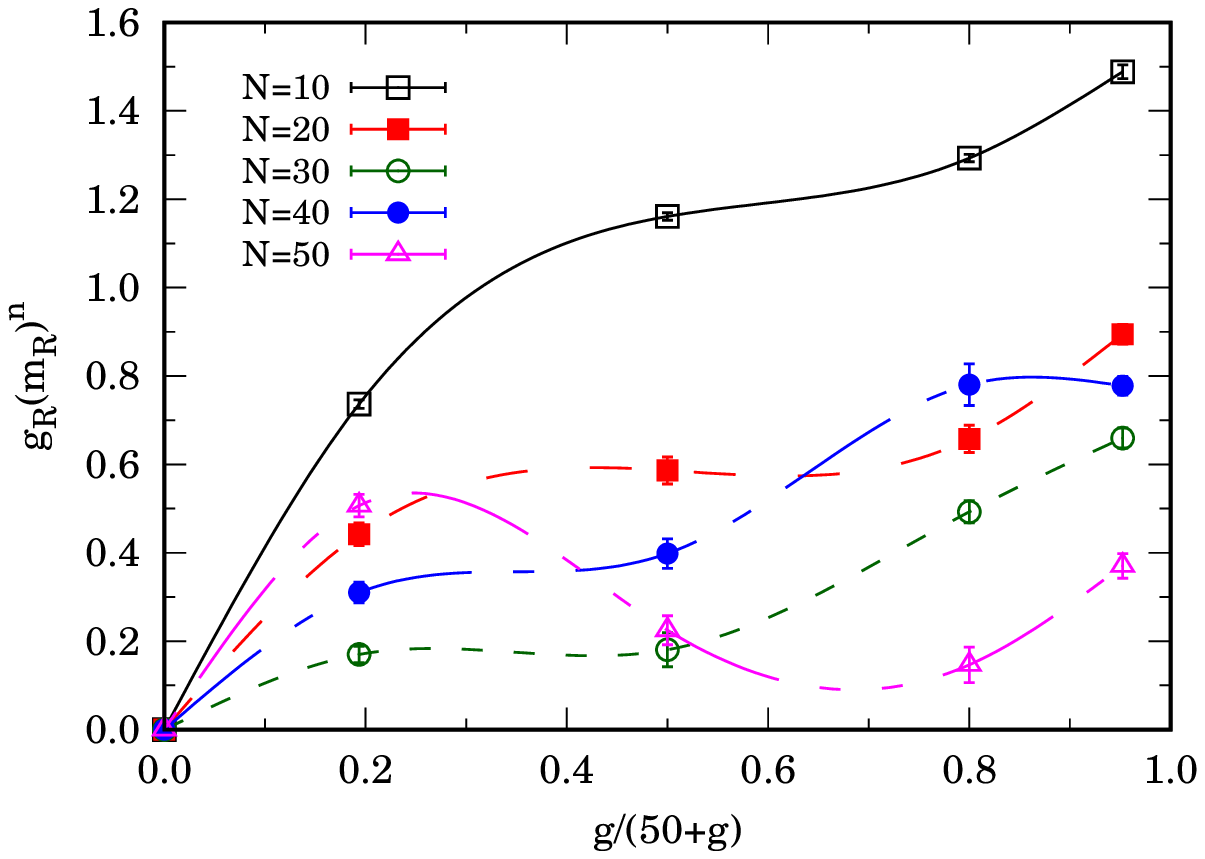}
\includegraphics[width=7cm]{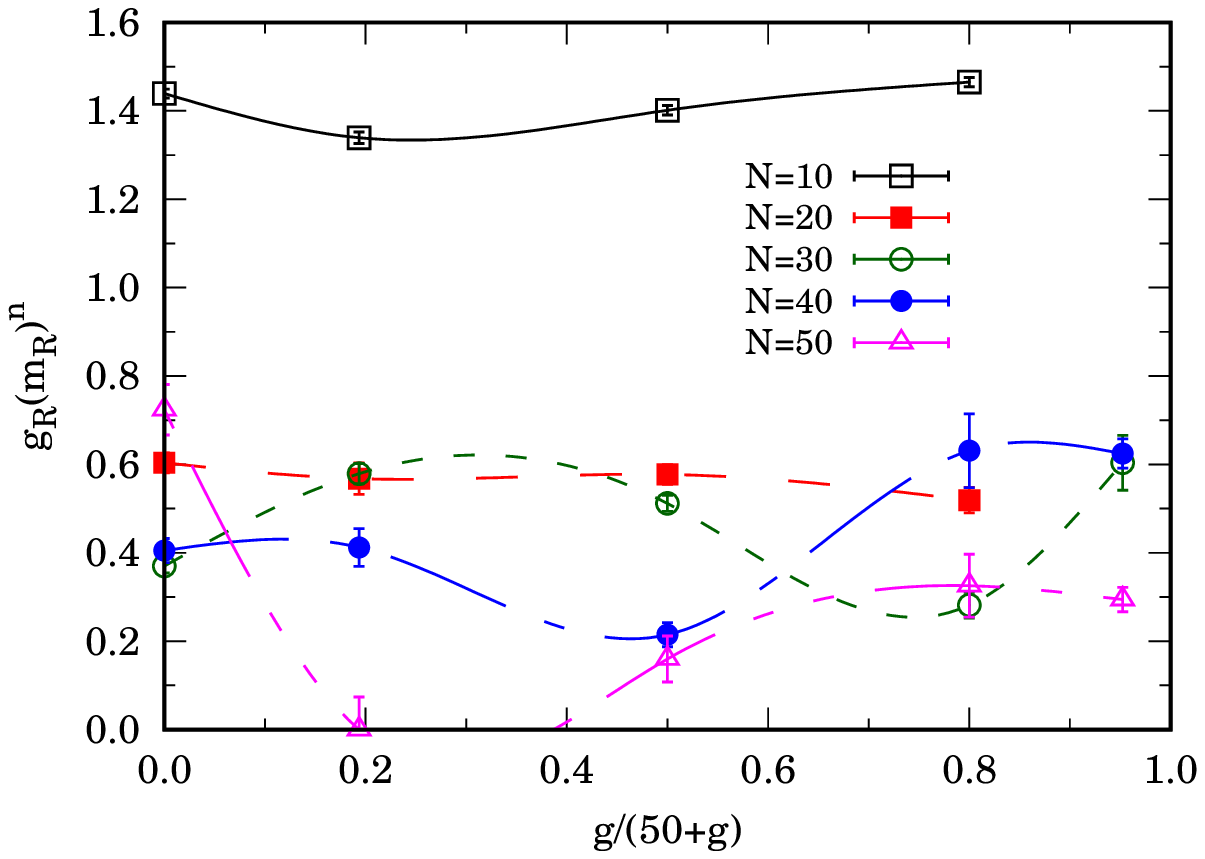}
\end{center}
\caption{(color online) For the scaled {\sl canonical} (left panels) and scaled {\sl affine} 
(right panels) $\vp^4_2$ ultralocal euclidean scalar field theory, we show the renormalized 
mass $m_R$ (top panels), the renormalized coupling constants $g_R$ (central panels), and 
$g_Rm_R^n$ (bottom panels) for various values of the bare coupling constant $g$ at decreasing 
values of the lattice spacing $a=1/N$ ($N\to\infty$ continuum limit). 
The statistical errors in the Monte Carlo were smaller than the symbols used. 
The main source of uncertainty is nonetheless the indirect one stemming from the unavoidable 
difficulty of keeping the renormalized mass constant throughout all cases. The 
lines connecting the simulation points are just a guide for the eye.}
\label{fig:a4-2}
\end{figure}

Following Freedman et al. \cite{Freedman1982}, for each $N$ and $g$, we adjusted the 
bare mass $m$ in such a way to maintain the renormalized mass approximately constant, 
$m_R\approx 3$, to within a few percent (in all cases less than $20\%$), and we measured the 
renormalized coupling constant $g_R$ for various values of the bare coupling constant $g$ at 
a given small value of the lattice spacing $a=1/N$ (this corresponds to choosing an absolute 
temperature $k_BT=1$ and a fixed length $L=1$). Note that in the CQ case it was necessary to 
choose imaginary bare masses for $g>0$. With $Na$ and $m_R$ fixed, 
as $a$ was made smaller, whatever change we found in $g_Rm_R^n$ as a function of $g$ 
could only be due to the change in $a$. We generally found that a depression in $m_R$ 
produced an elevation in the corresponding value of $g_R$ and viceversa; for this reason it 
is convenient to define an alternative renormalized coupling constant less sensitive to small 
variations of $m_R$, namely $g_R(m_R)^n$ (see \cite{Freedman1982}). The results are shown in 
Fig. \ref{fig:a4-2} for the scaled canonical action (\ref{eq:canonical-action}) and the 
scaled affine action (\ref{eq:scaled-affine-action}) in natural units $c=\hbar=k_B=1$, where, 
following Freedman et al. \cite{Freedman1982} we decided to compress the range of $g$ for 
display, by choosing the horizontal axis to be $g/(50+g)$. The constraint 
$m_R\approx 3$ was not easy to implement since for each $N$ and $g$ we had to run the 
simulation several times with different values of the bare mass $m$ in order to 
determine the value which would satisfy the constraint $m_R\approx 3$. This was the main 
source of uncontrolled uncertainty in the data.

In our simulations we used $10^8$ MC steps where in each step we attempt to move once all the 
$N^n$ fields variables of integration through the Metropolis algorithm \cite{Metropolis}. We 
used block averages and estimated the statistical errors using the jakknife method (described 
in Section 3.6 of \cite{Janke2002}) to take into account of the correlation time of the 
simulations. We always adjusted the field displacement in the random walk so to keep the 
acceptance ratios as close as possible to $1/2$.

Comparing the results for the scaled canonical and affine action we can see how
the renormalized coupling constant of the two approaches behaves very similarly at  
$g\neq 0$, \footnote{Comparing with the previous covariant studies 
\cite{Fantoni2020,Fantoni2020a,Fantoni2020b,Fantoni2021,Fantoni2021b,Fantoni2022,Fantoni2022b,Fantoni2022c} 
we can now say that removing the gradient term leads to a wilder behavior of the paths which 
could complicate finding any difference between CQ and AQ.} but in a neighborhood of $g=0$ 
the affine version remains far from zero in the continuum limit when the ultraviolet 
cutoff is removed ($Na=1$ and $N\to\infty$). The decrease of the renormalized coupling $g_R$
for increasing $N$ has to be expected, both for the CQ and the AQ cases, due to the use we 
made of the scaling $g\ra a^{s(r-2)/2} g$ which makes the model a ``free'' one in the 
continuum limit, $a\to 0$, when $r>2$. Of course the scaling we used has just a mathematical 
and not a physical justification (see footnote \ref{foot:math-phys}). In particular the 
scaling permits to have non diverging field expectation values in the AQ approach. Therefore 
as already observed in several previous covariant studies 
\cite{Fantoni2020,Fantoni2020a,Fantoni2020b,Fantoni2021,Fantoni2022b} we expect that also in 
this ultralocal case the AQ approach gives rise to a ``non-free'' field theory contrary 
to what happens for the CQ approach for $r>2$. This success of affine quantization to produce 
a well-defined, renormalizable, nontrivial, ``non-free'' quantum field theory is one of its 
merits and benefits.

During our simulations we kept under control also the vacuum expectation value of the 
field which in all cases was found to vanish in agreement with the fact that the 
symmetry $\vp\to-\vp$ of the scaled canonical action is preserved in the scaled affine case.
The random walk in the field is always able to tunnel through the barrier at $\varphi=0$ due 
to the affine effective term, $\frac{3}{8}(\hbar/\varphi)^2$, in the interaction. This is a 
consequence of working at finite $N$ and we expect the symmetry to be spontaneously broken in 
the continuum $N\to\infty$ limit 
\footnote{Once again this is only possible in a mathematical world but not in the physical 
(see footnote \ref{foot:math-phys}).}
when the point at $\varphi=0$ is excluded. Our results also 
show how the sum rules $g_R\to 0$ and $m_R\to m$ for $g\to 0$ are satisfied for CQ as it 
should for any gaussian weighting factor $\exp(-{\cal S}[\varphi])$ for any $N$.

%%%%%%%%%%%%%%%%%%%%%%%%%%%%%%%%%%%%%%%%%%%%%%%%%%%%%%%%%%%%%%%%%%%%%%%%%%%%%%
\section{Conclusions}
%%%%%%%%%%%%%%%%%%%%%%%%%%%%%%%%%%%%%%%%%%%%%%%%%%%%%%%%%%%%%%%%%%%%%%%%%%%%%%
In conclusion we performed a path integral Monte Carlo study of the properties (mass 
and coupling constant) of the renormalized ultralocal euclidean scalar field theory 
$\vp^4_2$ quantized through scaled affine and canonical quantization. Our results confirm the 
theoretical expectation for a ``free'' theory in the continuum limit. This is merely a 
consequence of the chosen scaling. As in previous works on covariant theories we expect that 
also in this ultralocal case the un-scaled AQ approach gives rise to a ``non-free'' field 
theory contrary to what happens for the un-scaled CQ approach for $r>2$. Indeed already for 
the scaled version in the AQ theory the renormalized coupling does not seem to go towards 
zero at least when the bare coupling is zero when one approaches the continuum limit, as 
stems from our path integral Monte Carlo results. This means that a ``free'' scaled AQ theory 
is profoundly different from a ``free'' scaled CQ one: the former is therefore non-trivial 
and renormalizable and the latter is trivial and non-renormalizable.

%%%%%%%%%%%%%%%%%%%%%%%%%%%%%%%%%%%%%%%%%%%%%%%%%%%%%%%%%%%%%%%%%%%%%%%%%%%%%%
%\bibliographystyle{}
\bibliography{four-four}

%%%%%%%%%%%%%%%%%%%%%%%%%%%%%%%%%%%%%%%%%%%%%%%%%%%%%%%%%%%%%%%%%%%%%%%%%%%%%%

%%%%%%%%%%%%%%%%%%%%%%%%%%%%%%%%%%%%%%%%%%%%%%%%%%%%%%%%%%%%%%%%%%%%%%%%%%%%%%
%%%%%%%%%%%%%%%%%%%%%%%%%%%%%%%%%%%%%%%%%%%%%%%%%%%%%%%%%%%%%%%%%%%%%%%%%%%%%%
%%%%%%%%%%%%%%%%%%%%%%%%%%%%%%%%%%%%%%%%%%%%%%%%%%%%%%%%%%%%%%%%%%%%%%%%%%%%%%
\end{document}